\documentclass[
 bmf,
 amsmath,
 amssymb,
 preprint,%
]{revtex4-1}

\usepackage{xcolor}
\usepackage{graphicx}
\usepackage{dcolumn}
\usepackage{bm}
\usepackage[mathlines]{lineno}

\usepackage[utf8]{inputenc}
\usepackage[T1]{fontenc}
\usepackage{mathptmx}
\usepackage{etoolbox}

\usepackage{soul}

\makeatletter
\def\@email#1#2{%
 \endgroup
 \patchcmd{\titleblock@produce}
  {\frontmatter@RRAPformat}
  {\frontmatter@RRAPformat{\produce@RRAP{*#1\href{mailto:#2}{#2}}}\frontmatter@RRAPformat}
  {}{}
}%
\makeatother
\begin{document}

\title[]{Synchronous oscillatory electro-inertial focusing of microparticles}

\author{Giridar Vishwanathan}
\author{Gabriel Juarez}
 \email{gjuarez@illinois.edu.}
\affiliation{Department of Mechanical Science and Engineering, University of Illinois Urbana-Champaign, Urbana, IL 61801, USA}

\date{\today}

\begin{abstract}

Here, results are presented on the focusing of $1 \ \mu$m polystyrene particle suspensions using a synchronous oscillatory pressure-driven flow and oscillatory electric field in a microfluidic device. 
The effect of the phase difference between the oscillatory fields on the focusing position and focusing efficiency was investigated. 
The focusing position of negatively charged polystyrene particles could be tuned anywhere between the channel centerline to the channel walls. 
Similarly, the focusing efficiency could range from 20\% up to 90\%, depending on the phase difference, for particle Reynolds numbers of order $O(10^{-4})$.
The migration velocity profile was measured and the peak velocity was found to scale linearly with both the oscillatory pressure-driven flow amplitude and oscillatory electric field amplitude.
Furthermore, the average migration velocity was observed to scale with the cosine of the phase difference between the fields, indicating the coupled non-linear nature of the phenomenon. 
Lastly, the peak migration velocity was measured for different particle radii and found to have an inverse relation, where the velocity increased with decreasing particle radius for identical conditions.



\end{abstract}

\maketitle

\section{Introduction}

The applications of electrokinetic effects in microfluidics are widespread and well-studied.
Some prominent examples of these applications are the use of electro-osmosis for microfluidic pumping \cite{Li2012},  electrophoresis for particle separation through electrical flow-field fractionation \cite{waheed}, electro-wetting on dielectric (EWOD) for droplet control \cite{Rui}, dielectrophoresis for cell manipulation \cite{Pethig}, and AC electro-osmotic and AC electro-thermal vortices for mixing and bioparticle separation \cite{Ng,Qi2016}. 
All of these applications utilize electric fields but differ primarily by the frequency of the applied electric field and the configuration of the electrodes with respect to the fluid flow. 
The electric field frequencies range from $f=0$ Hz for DC applications such as EWOD and electro-osmotic pumping, $1 < f < 100$ KHz for AC electro-osmotic flow  $0.1 < f < 10$ MHz for dielectrophoresis and AC electrothermal flow.

Likewise, the applications of hydrodynamic inertia in microfluidics are also well explored.
Hydrodynamic inertial forces are highly effective in separating $3 < a < 30\ \mu$m particles from suspensions of smaller particles in a simple, clog-free, and high-throughput manner. Such a technique is highly useful for the processing of biomedical samples with the bioparticles of interest being cancer cells, platelets or bacterial cells found in blood samples \cite{DiCarlo2009}. 
Most inertial microfluidics techniques employ a steady flow and vary depending on the channel geometry, such as straight, serpentine, spiral or contraction-expansion \cite{Gangadhar,Jiang,Martel2}. 
The use of oscillatory flows for inertial microfluidics is more recent and has the potential to manipulate sub-micron particles \cite{Mutlu}. 
Attempts towards quantifying the role of oscillation frequency \cite{VishwanathanJFM} and channel curvature \cite{BhosalePNAS} have also been made.

Recently there is a growing interest in investigating the non-linear interaction between inertial and electro-kinetic phenomena. 
It has been found that negatively charged particles suspended in a liquid that is simultaneously subjected to flow and a parallel DC electric field undergo migration to the center of the channel cross-section \cite{KimYoo, Liang}. 
When the direction of the electric field is reversed, or anti-parallel, to the flow the particles migrate towards the walls \cite{Yuan}. 
The direction of migration was also found to be exactly reversed when a viscoelastic solution is used instead of a Newtonian one \cite{Li}.
This phenomenon has the potential to combine the high throughput of inertial methods and the precise control of electrokinetic methods.

In this work, we investigate the migration of suspended microparticles simultaneously subjected to an oscillatory electric field and a synchronized oscillatory pressure-driven flow, or synchronous oscillatory electro-inertial focusing (SOEIF), at a frequency of 50 Hz. 
This manuscript is organized as follows. 
First, the concept of SOEIF and the experimental setup used to realize it is described. 
Then, particle focusing of polystyrene particles of radius $0.5 \ \mu$m is experimentally demonstrated. 
Next, the steady-state focusing position and focusing efficiency as a function of the phase difference between the oscillatory pressure-driven flow and the oscillatory electric field is characterized. 
Then, the migration velocity profile for different values of the phase difference is measured and compared.
Finally, the migration velocity for different particle sizes at identical flow and field conditions is measured and the scaling of migration velocity with particle radius is reported.


\section{Problem formulation}

\begin{figure}
\centering
\includegraphics[width=\linewidth]{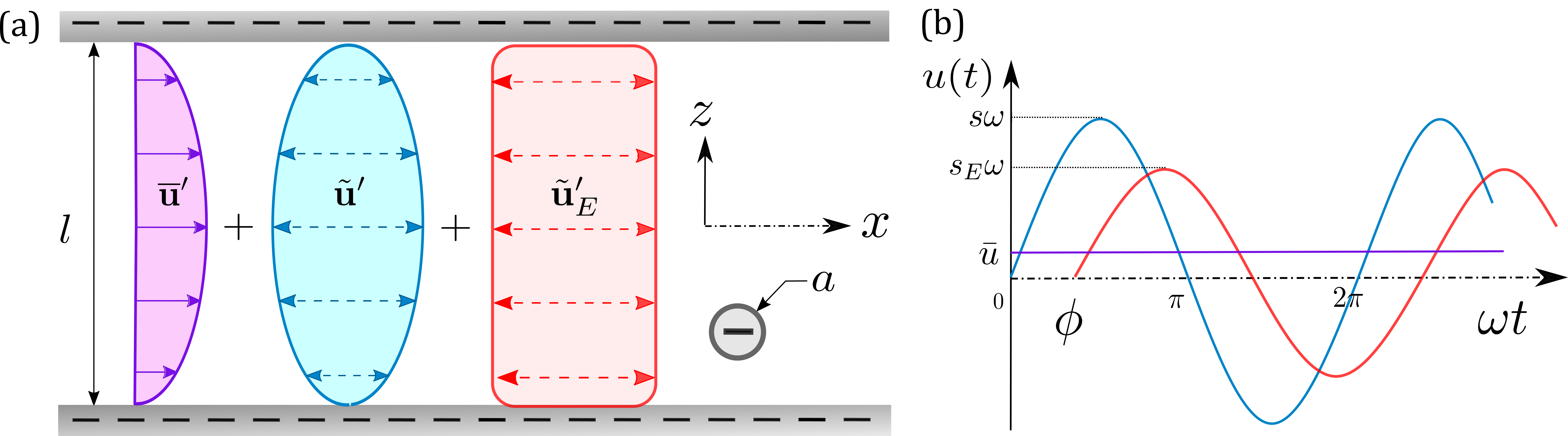}
\caption{ 
(a) Idealized illustration of synchronous oscillatory electro-inertial focusing (SOEIF) of an insulating, negatively charged, neutrally buoyant spherical particle in an insulating, negatively charged 2D channel.
(b) Illustration of the centerline velocities versus time for each oscillatory flow component.
}
\label{fig:figonesoeif}
\end{figure}

Consider the idealized configuration illustrated in Figure \ref{fig:figonesoeif}(a). 
A rigid, neutrally buoyant, insulating, negatively charged spherical particle of radius $a$ and zeta potential $\zeta_p$ is suspended in a Newtonian liquid of kinematic viscosity $\nu$, ionic conductivity $\sigma$, permittivity $\varepsilon$ as it flows through a 2D insulating channel with width $l$ and zeta potential $\zeta_w$. 
The underlying flow in the channel consists of a steady flow component $\bar{\mathbf{u}}'(z)$, a pressure-driven oscillatory flow component $\tilde{\mathbf{u}}'(z,t')$ and an electro-osmotic oscillatory flow component $\tilde{\mathbf{u}}'_E(z,t')$ that is caused by a uniform oscillating electric field $\tilde{\mathbf{E}}'(t)$.
The temporal variation of the three velocity components at $z=0$ is illustrated in Figure \ref{fig:figonesoeif}(b). 
Both the pressure-driven flow and the electro-osmotic flow have the same frequency $f$ and angular frequency $\omega=2\pi f$. 
The phase difference between the electro-osmotic velocity and the oscillatory flow is given by $\phi$. 
The amplitude of the centerline pressure-driven and electro-osmotic flow are $U$ and $U_E$ respectively. 
The amplitude of the electric field is $E$ and the centerline steady flow velocity is $\bar{u}$.

\section{Experimental methods}

\begin{figure}
	\centering
\includegraphics[width=\linewidth]{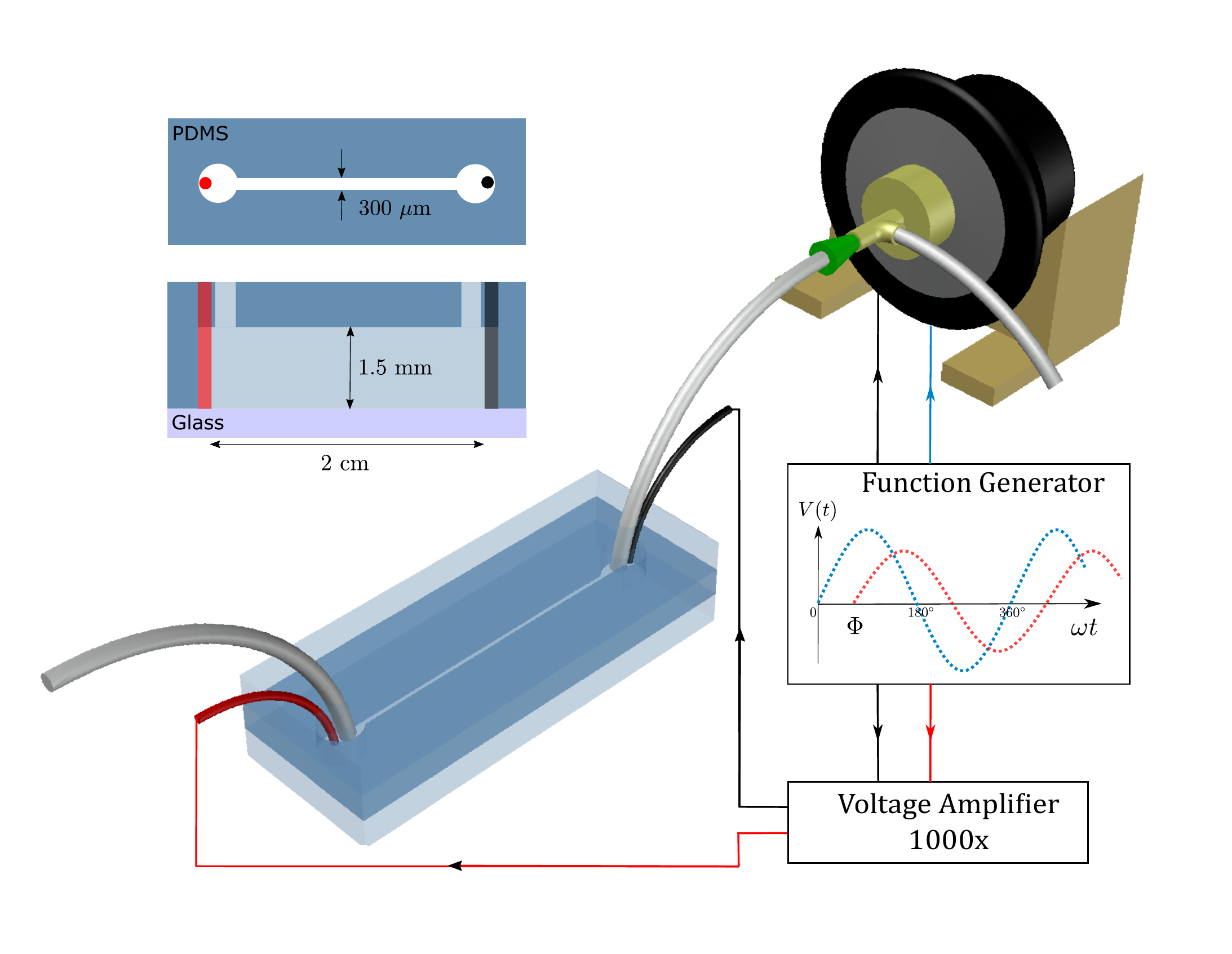}
\caption{
Schematic of the experimental setup and microchannel. 
The PDMS microchannel (top left) has length $L=2$ cm, height $h=1.5$ mm and width $l=300 \ \mu$m. 
A polyethylene tube and a stainless steel electrode is inserted at both the channel inlet and outlet. 
The outlet tube is interfaced to a speaker using an adapter elbow (yellow) affixed to the speaker cone. 
A pipette tip (green) is used as a wedge seal. 
The electrode terminals are connected to a high-voltage adapter with a $1000\times$ gain. 
A function generator activates the amplifier and the speaker, producing synchronized sinusoidal voltages with a controlled phase difference $\Phi$.
}
	\label{fig:figtwosoeif}
\end{figure}


A schematic of the experimental setup is shown in Figure \ref{fig:figtwosoeif}. 
The flow manifold is a straight rectangular microchannel made of PDMS (10:1, resin:crosslinker), de-molded from 3D printed mold (SLA, Form3) treated with silicone mold release. 
The microchannel has length $L=2$ cm, height $h=1.5$ mm and width $l=300 \ \mu$m. 
Four holes are made with a biopsy punch (1mm), two at each end of the channel. 
The channel is then bonded to a glass slide after 1 minute of oxygen plasma treatment. 
After bonding, a few drops of uncured PDMS are poured along the edges of the PDMS-glass interface and subsequently cured for better bonding and sealing. 
Two multi-stranded copper wires with stainless steel pins are used for the electrodes. 
Polyethylene tubing is inserted at the inlet and outlet, followed by the pins to create the electrodes. 
The tubing and electrode interfaces are sealed with epoxy (Loctite 5 min).


To produce the flow, a speaker-based oscillatory source capable of producing $s\geq100\ \mu$m and $U\leq8$ cm/s across the frequency range $25\leq f \leq 800$ Hz in the PDMS microchannels \cite{VishwanathanMicroNano} is used along with a syringe pump operated at $10 - 20\  \mu$L/min, that produces a steady flow with a centerline speed $\bar{u}=1-2$ mm/s. 
A high voltage amplifier (TREK 677) with a gain of $1000\times$ is used to generate an oscillatory electric field. 
The high voltage line is supplied from the amplifier to two alligator clips that then grip the copper strands of either electrode. 
The sinusoidal input voltages ($V(t)$ with $0 - 5$ Vpp) to both the speaker and amplifier are supplied by a function generator (RIGOL DG812) with a controlled phase difference $\Phi$. 
The majority of experiments are performed at $f=50$ Hz with an electric field amplitude $E=50$ kV/m, which corresponds to 2000 Vpp (peak-to-peak voltage) applied across a gap of 2 cm.



The dynamics of the particles under the combined influence of the electric field and pressure-driven flow are visualized with brightfield imaging using an inverted microscope (Nikon Eclipse Ti2). 
For measuring particle focusing, a lens with $4\times$ magnification is used to image particles near the channel outlet. 
For measuring migration velocity, a lens with $10 \times$ magnification is used to image particles near the channel inlet. 
All data is obtained at the channel midplane.


The solution used was $22\%$ glycerol in water with $0.1\ $mM NaCl, which corresponds to a kinematic viscosity $\nu=1.68\ \mathrm{mm^2/s}$, density $\rho=1060\ \mathrm{kg/m^3}$ \cite{Glycerine}, conductivity $\sigma=0.12$ mS/cm \cite{Liao}, permittivity $\varepsilon=6.55\times10^{-10}$ F/m \cite{Behrends} and effective ionic diffusion coefficient $D=1.15\times10^{-9} \ \mathrm{m^2/s}$.
Dilute particle suspensions were made using polystyrene particles (Spherotech) with radii (and volume fractions) of $a=0.5 \ \mu$m $(0.001\%)$, $1.7 \ \mu$m $(0.003\%)$, or $2.6\ \mu$m $(0.01\%)$.
The laboratory temperature is maintained at $20 \ ^{\circ}$C.


For the parameter ranges used here, velocity magnitudes are such that: $\bar{u}, U_E \ll U$ implying that purely hydrodynamic effects are largely determined by the pressure-driven oscillatory flow. 
The length dimensions are likewise related by $a\ll l\ll h \ll L$. 
This implies that flow inside the channel is fully developed and the flow is quasi 2D at the midplane, where data is obtained.

The channel Reynolds number is given by $\mathrm{Re}=U l/\nu$ and for all experiments considered here is $\mathrm{Re}\leq10$, which implies that the flow profile is unaltered by inertial effects. 
The Womersley number $\alpha=l\sqrt{\omega/\nu}$, governs transient hydrodynamic effects at the channel scale and therefore determines $\tilde{\mathbf{u}}'(z,t)$ and $\tilde{\mathbf{u}}'_E(z,t)$. 
For a frequency of $50$ Hz, the Womersley number is $\alpha=4.1$, which corresponds to a quasi-steady flow profile \cite{Obrien}. 
Taken together, the ranges of $\mathrm{Re}$ and $\alpha$ imply that for $-l/2+O(\lambda_D) < z < l/2-O(\lambda_D)$, the pressure driven flow is given by:
\begin{equation}
	\tilde{\mathbf{u}}'(z,t)\approxeq U(1-(2z/l)^2)\sin{(\omega t)},\label{uoscisoeif}
\end{equation} 
and the electro-osmotic flow is given by :
\begin{equation}
\tilde{\mathbf{u}}_E'(z,t)\approxeq U_E\sin{(\omega t+\phi)}.\label{ueofsoeif}
\end{equation}

The Stokes number $St=2a^2\omega/9\nu$ governs transient hydrodynamic effects at the scale of the particle. 
For the experimental parameters in this work, the Stokes number is $St\leq3\times10^{-4}$, which is $St \ll 1$ and implies that transient effects at the particle scale are irrelevant \cite{Michaelides}. 
Similarly, the particle Reynolds number $\mathrm{Re_p}=U a^2/l \nu$, which governs inertial effects at the particle scale, takes values in the range $10^{-4} < \mathrm{Re_p} < 10^{-3}$. 
For this range, forces responsible for inertial migration are weaker than Brownian motion \cite{Mutlu}.


The principal electrokinetic effects relevant to these experiments are electro-osmotic flow at the channel scale and electrophoresis at the particle scale. 
The amplitude of the electro-osmotic flow in terms of electric field is $\varepsilon \zeta_w E/\mu$. 
Likewise, the electrophoretic slip velocity amplitude at the surface of the particle is $\varepsilon \zeta_p E/\mu$. 
Here, $\zeta_w$ and $\zeta_p$ are the zeta potentials of the wall and the particle, respectively.

For the low frequency ($\alpha < 5$) experiments performed herein, it is not possible to separate the effects of electrophoresis and electro-osmosis because the electro-osmotic flow can only be visualized using tracer particles. 
For this reason, for each experiment, the amplitude of the combined measurement is reported as:
\begin{equation}
	s_E\omega=\varepsilon(\zeta_p+\zeta_w)E/\mu \ .
\end{equation} 

However, at high frequency ($\alpha \gg 5$), the electro-osmotic flow component is suppressed near the channel center. 
This principle was used to measure the particle zeta potentials by direct visualization in the microchannel using fast single-particle tracking \cite{Cid}.
For this, the oscillatory motion of particles near the channel center for an electric field amplitude of $E=125$ kV/m, and frequency of $800$ Hz ($\alpha=16.5$) was tracked using a global-shutter high-speed camera recording at 16000 fps. 
For particle radii $a=0.5,\ 1.7,\ 2.6\ \mu$m, the particle zeta potentials computed were $\zeta_p=-22\pm2, -39\pm8,$ and $-53\pm5$ mV, respectively.  

The electro-osmotic flow velocity amplitude can therefore be obtained by subtracting the electrophoretic component. 
The ratio of the electric potential energy to the thermal potential energy ranges $0.8 \leq e\zeta_p/k_BT \leq 2.0$. 
For this range, the nonlinear electrokinetic effects are weak \cite{Dutta}. 
The ratio of viscous to electrokinetic stresses is estimated by $Q=sa/s_El<0.2$ for the experimental parameter range accessed here. 

Finally, the colloidal timescales of significance are: the Maxwell-Wagner timescale $\tau_{MW}=\varepsilon/\sigma\approxeq65$ ns, the volume diffusion/concentration polarization timescale $\tau_{VD}=a^2/2D\approxeq0.2$ ms \cite{Shilov}, the double layer polarization timescale $\tau= \lambda^2_D/D\approxeq0.2 \ \mu$s.
Here, $\lambda_D$ is the Debye length $\lambda_D = \sqrt{\frac{\varepsilon k_BT}{2z^2e^2C_0}}$, where $k_BT$ is the product of the Boltzmann constant and temperature, $z$ is the valence of the ionic species, $e$ is the electronic charge magnitude, and $C_0$ is the ionic number density.
For the experiments, the Debye length is $\lambda_D \approxeq 16.8$ nm.
All of these timescales are significantly smaller than the oscillatory time period of $20$ ms for an oscillatory frequency of 50 Hz, which implies quasi-steady electrokinetic behavior at the particle scale.

\subsection*{Phase difference: $\Phi$ versus $\phi$} 

It is worth highlighting the difference between $\Phi$ and $\phi$. 
The phase difference set by the function generator between the voltage applied to activate the electric field and the voltage applied to the speaker is $\Phi$ and can be experimentally controlled. 
The actual phase difference between the electro-osmotic oscillatory flow and the pressure-driven oscillatory flows is $\phi$ and it cannot be controlled experimentally. 
All subsequent results are reported in terms of $\Phi$ only, the relationship between $\Phi$ and $\phi$ is elaborated below.

The principal relation between $\phi$ and $\Phi$ is $\phi=\Phi+\phi_{s}+\phi_{sE}$, where $\phi_{s}$ is a phase shift associated with the startup of the pressure-driven flow and $\phi_{sE}$ the phase shift associated with the startup of the electro-osmotic flow.
The electric field, is exactly synchronized with the applied voltage and there is no measurable phase lag for the frequencies considered here, that is $\phi_{sE}\approx0$. 
The high voltage output of the amplifier was verified to be synchronized with the voltage output of the function generator correct to $1 \times 10^{-6}$ s. 
Further, since the channel materials: PDMS, glass and polyethylene are completely insulating, their conductance and inductance are negligible. 
The capacitive and resistive impedances of the fluid in the channel can be estimated by $X_c=L/\varepsilon l h\omega = 2\times10^{11}\ \Omega$ and $R=L/\sigma l h=4.4\times10^{6}\ \Omega$ indicating that load is almost entirely resistive. 
This implies that the current and electric field inside the channel are exactly synchronized with the voltage. 
It follows that the electro-osmotic flow velocity and the electrophoretic particle migration is anti-synchronized owing to the negative charges of the channel wall and particle.

Likewise, the input voltage to the speaker is measured to be synchronized with the output of the function generator. 
The relatively large wavelengths of the frequencies used here ($\lambda=c/f=30$ m) imply that the phase shift effects within the fluid and other mechanical components such as the tubes, and adapter are also negligible. 
Nevertheless, the effects of $\phi_s$ are not necessarily negligible. 
This is because all uncertainty about the relationship between $\Phi$ and $\phi$ arises during the startup of the speaker. 
The phase characteristics of the speaker cone excursion as a function of input voltage are not known and this evaluation is quite involved and therefore not performed here. 
To minimize the uncertainty associated with the speaker startup, the speaker activation protocol is standardized. 
Before each dataset, the speaker voltage is set to ``0'' and the cone is allowed to reach the equilibrium, or null position. 
Then the desired speaker voltage is set in the function generator and the speaker is activated from rest, after which it is kept undisturbed. 
Only the field voltage is changed to measure the effects of the phase. 
Furthermore, for each experiment, the phases difference are uniformly sampled to minimize any uncertainty associated with a finite $\phi_{s}$.


\section{Results}

\begin{figure}
\centering
\includegraphics[trim=0 150 120 0,clip,width=\linewidth]{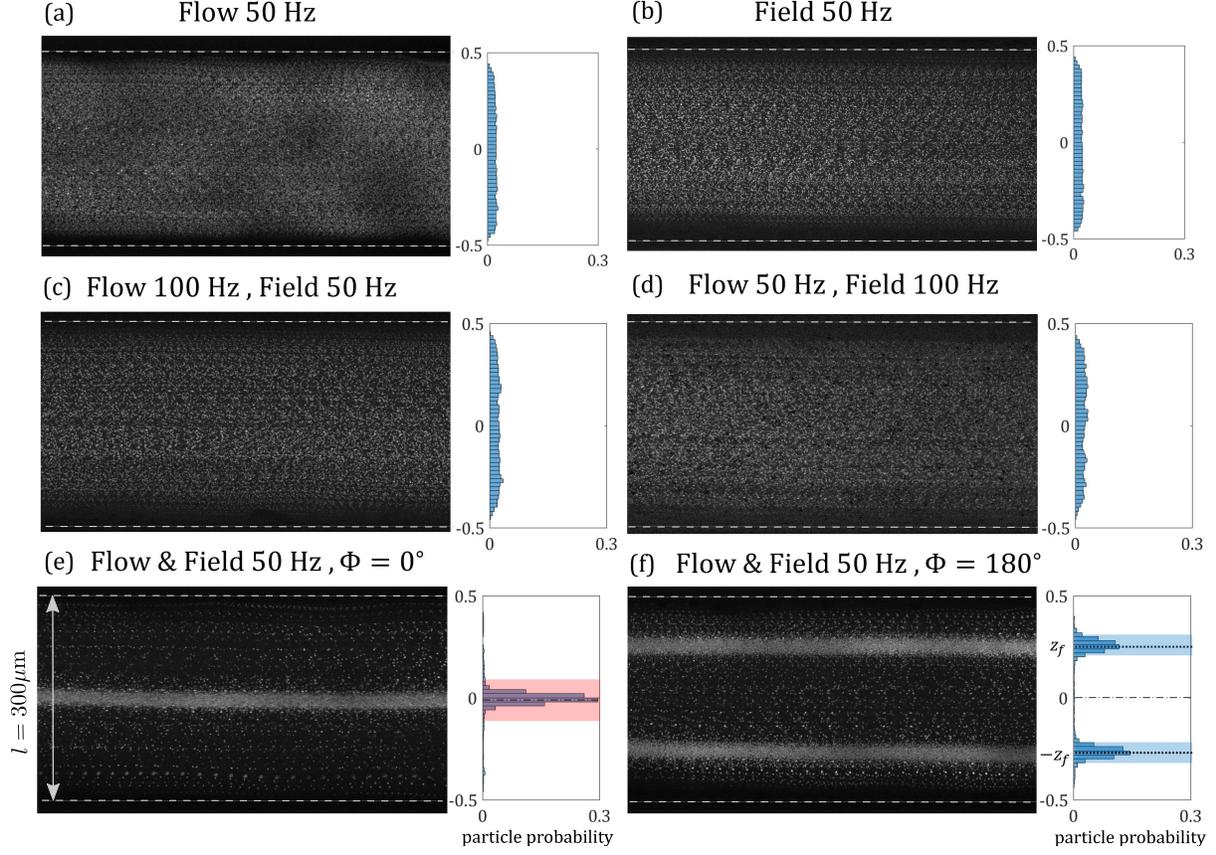}
\caption{
Particle pathlines (bright) and corresponding particle density histograms near the channel outlet for a suspension of $a=0.5\ \mu$m particles, a transport flow of $\bar{u}=2$ mm/s, and different asynchronous and synchronous flow conditions. 
(a) Oscillatory pressure-driven flow ($f=50$ Hz, $U=73\pm5$ mm/s, $E=0$ kV/m, $s_E\omega=0$ mm/s) produces a uniform particle distribution. 
(b) Oscillatory electric field ($f=50$ Hz, $U=0$ mm/s, $E=50$ kV/m, $s_E\omega=1.1\pm0.2$ mm/s) produces a uniform particle distribution. 
(c) Asynchronous oscillatory pressure-driven flow ($f=100$ Hz, $U=77\pm8$ mm/s) and oscillatory electric field ($f=50$ Hz, $E=50$ kV/m, $s_E\omega=1.1\pm0.2$ mm/s) produces a uniform particle distribution.
(d) Asynchronous oscillatory pressure-driven flow ($f=50$ Hz, $U=73\pm5$ mm/s) and oscillatory electric field ($f=100$ Hz, $E=50$ kV/m, $s_E\omega=0.9\pm0.2$ mm/s) produces a uniform particle distribution. 
(e) Synchronous oscillatory pressure-driven flow ($f=50$ Hz, $U=73\pm5$ mm/s) and oscillatory electric field ($f=50$ Hz, $E=50$ kV/m, $s_E\omega=1.1\pm0.2$ mm/s) with $\Phi=0^{\circ}$ produces a single focusing band at the channel center. 
(f) Synchronous oscillatory pressure-driven flow ($f=50$ Hz, $U=73\pm 5$ mm/s) and oscillatory electric field ($f=50$ Hz, $E=50$ kV/m, $s_E\omega=1.1\pm 0.2$ mm/s) with $\Phi=180^{\circ}$ produces two focusing bands near the channel walls.
}
\label{fig:figthreesoeif}
\end{figure}

The concept of SOEIF is demonstrated for a suspension of particles with radii $a=0.5\ \mu$m  transported by a steady flow with a speed $\bar{u}=2$ mm/s from left to right within a microfluidic channel, as shown in Figure \ref{fig:figthreesoeif}.
The experimental conditions for each image are listed in the caption.
These different conditions illustrate particle distribution under oscillatory pressure-driven flow, oscillatory electric field, and asynchronous or synchronous oscillatory electro-inertial flows. 
Example images are generated by subtracting the median background, and subsequently projecting 200 of the resulting frames (corresponding to 4 seconds) onto a single image. 
The bright dots correspond to tracer particles and the bright bands to regions of high particle density.

The distribution of particles across the channel cross-section is homogeneous when a transport flow is combined with only an oscillatory pressure-driven flow or only an oscillatory electric field, as shown in Figure \ref{fig:figthreesoeif}(a) and Figure \ref{fig:figthreesoeif}(b), respectively.
A similar homogeneous distribution is obtained when the oscillatory pressure-driven flow and electric field frequencies are not synchronized, as shown in Figure \ref{fig:figthreesoeif}(c) and Figure \ref{fig:figthreesoeif}(d). 
When the oscillatory pressure-driven flow and electric field are synchronized, however, the particles are observed to migrate perpendicular to the streamlines into highly localized focusing regions. 
Interestingly, the number of focusing bands and the focusing position are observed to vary depending on the phase difference between the oscillatory flow and the electric field.
For $\Phi=0^{\circ}$, there is a single focusing band at the channel centerline, as shown in \ref{fig:figthreesoeif}(e). 
Conversely, for $\Phi=180^{\circ}$, two focusing bands are symmetric about the channel centerline, as shown in Figure \ref{fig:figthreesoeif}(f).

Histograms of particle probability along the span-wise channel position are computed using particle identification and tracking velocimetry. 
The histograms are shown beside the corresponding projection images.
For the first four conditions, histograms confirm that suspended particles are uniformly distributed across the $300\ \mu$m channel width, shown in Figure \ref{fig:figthreesoeif}(a-d).
For synchronous oscillatory flow and field at $\Phi=0^{\circ}$, there is a single sharp peak in the histogram that corresponds to particles focusing at the channel centerline, as shown in Figure \ref{fig:figthreesoeif}(e).
Similarly, for synchronous oscillatory flow and field at $\Phi=180^{\circ}$, there are two peaks in the histogram that corresponds to the two bands that are symmetric about the channel centerline, as shown in Figure \ref{fig:figthreesoeif}(f).

The focusing position $z_f$ and the focusing efficiency $F_{20\%}$ were also obtained using particle tracking velocimetry. 
The channel centerline is set to be $z = 0$.
The focusing position $z_f$ is then defined by the distance of a histogram peak location to the channel centerline.
This measurement is sensitive to the position of the centerline and the width of the channel.
Therefore, particle tracking was used to extract the steady transport flow profile that was then fit to a parabolic curve, which corresponds to viscous flow.
The fitting constants determined the precise centerline and local channel width.
The focusing efficiency $F_{20\%}$ is defined as the percentage of total particles within a finite region surrounding the focusing position, or histogram peak. 
For example, if there is a single central focusing position, then the focusing efficiency is the percentage of total particles within $\pm l/10$ of the center histogram peak, indicated by the red shaded region, as shown in Figure \ref{fig:figthreesoeif}(e).
If there are two focusing positions symmetric about the centerline, then the focusing efficiency is the total fraction of particles located within $\pm l/20$ of either histogram peak, as shown by the blue shaded region in Figure \ref{fig:figthreesoeif}(f).
These metrics will be used to characterize the effect of phase difference $\Phi$ on the focusing position and focusing efficiency of a particle suspension in synchronous oscillatory electro-inertial flow.

\subsection{Focusing position and focusing efficiency}

\begin{figure}
\centering
\includegraphics[trim=30 0 30 0,clip,width=\linewidth]{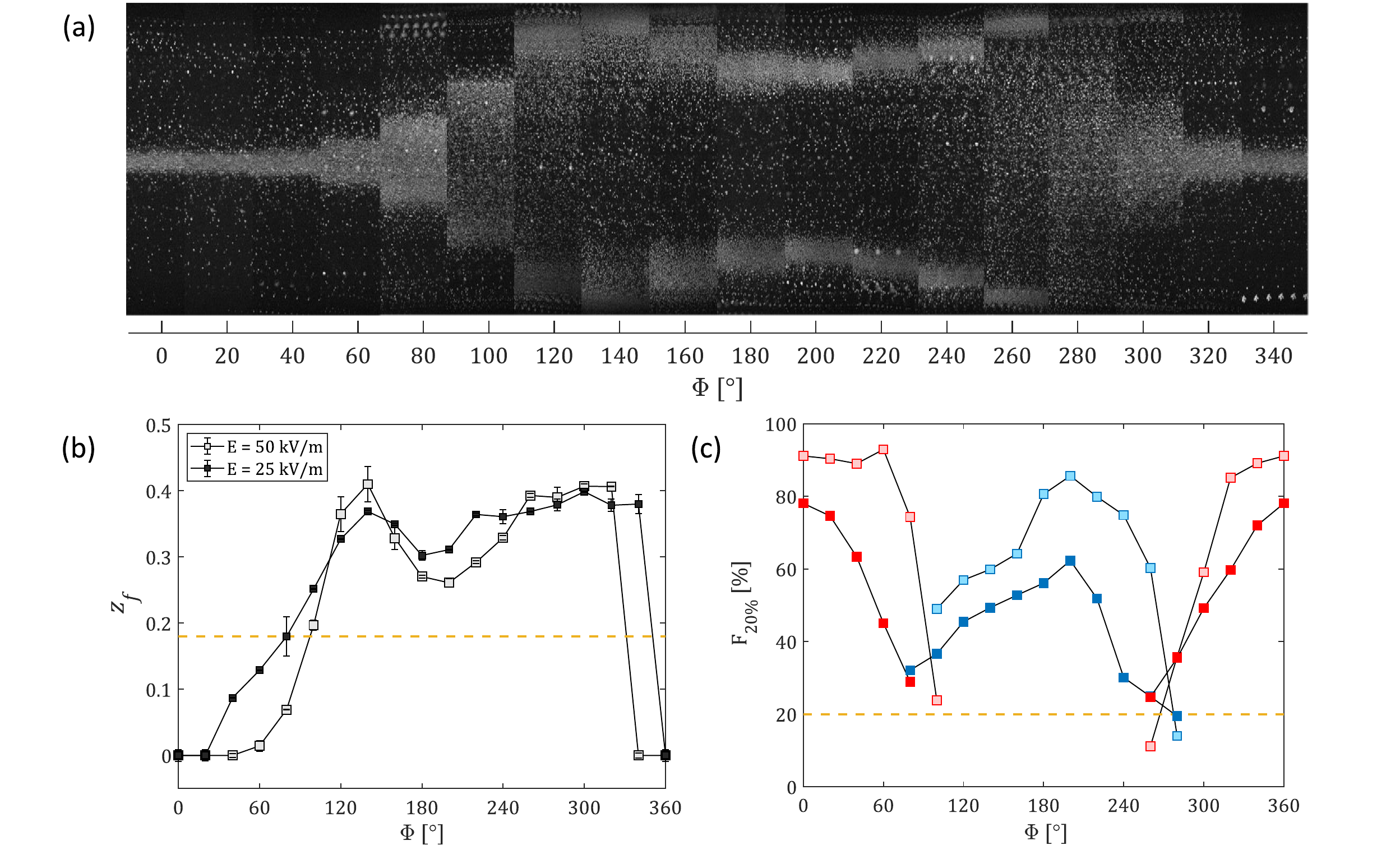}
\caption{
(a) Synchronous oscillatory electro-inertial focusing bands for $E=50$ kV/m as a function of the phase difference $\Phi$.
(b) Focusing position $z_f$ for $E=50$ kV/m (light) and $E=25$ kV/m (dark) as a function of the phase difference $\Phi$. 
The yellow dashed line indicates the focusing position for purely inertial focusing.
(c) Focusing efficiency $F_{20\%}$ for $E=50$ kV/m (light) and $E=25$ kV/m (dark) as a function of the phase difference $\Phi$. 
The $F_{20\%}$ refers to the percentage of total particles within $\pm\ l/10$ of the focusing position. 
The yellow dashed line indicates the efficiency of a uniform particle distribution.
Relevant experimental parameters are: $a=0.5\ \mu$m, $\bar{u}=2$ mm/s, $f=50$ Hz, $s=234\ \mu$m, and $s_E=3\ \mu$m.
}
\label{fig:figfoursoeif}
\end{figure}

The evolution of particle pathlines with the phase difference $\Phi$ shows a surprisingly rich behavior, as shown in Figure \ref{fig:figfoursoeif}(a). 
At $\Phi=0^{\circ}$ the particles focus into a single band at the channel centerline. 
For $0^{\circ} < \Phi < 40^{\circ}$, the single band focusing persists with the same approximate band thickness. 
For $60^{\circ} < \Phi < 80^{\circ}$, the apparent thickness of the focusing band increases in width. 
For $100^{\circ} < \Phi < 240^{\circ}$ there are two distinct, well-separated focusing bands symmetric about the channel centerline. 
There is also no observable focusing at the channel center. 
At $\Phi=260^{\circ}$, there is a resurgence of an observable but diffuse focusing band at the center, which coexists with the two focusing positions close to the channel walls. 
As $\Phi$ increases further to $\Phi=300^{\circ}$, the focusing band at the center becomes narrower and more well-defined, while the focusing band near the walls becomes less defined in comparison. 
For $\Phi \geq 320^{\circ}$, the focusing bands near the walls become unresolvable, while the focusing band at the center continues to become narrower and more distinct, returning to the case of $\Phi=0^{\circ}$.

The evolution of focusing position $z_f$ is plotted against phase difference, as shown in Figure \ref{fig:figfoursoeif}(b). 
Here, the channel centerline corresponds to $z_f=0$ and the channel walls correspond to $z_f=0.5$.
The yellow dashed line is $z_f=0.18$, which represents the expected focusing position that would be due to solely inertial focusing \cite{Ho,VishwanathanJFM}.
Initially, for $0\leq\Phi\leq20^{\circ}$, the focusing position is at the channel centerline so that $z_f\approx0$. 
For $40^\circ\leq\Phi\leq80^\circ$, the centerline band widens as the transition from single band focusing to two band focusing occurs.  
For $100^{\circ} \leq \Phi \leq 260^{\circ}$, the two band focusing positions move toward the channel walls and display non-monotonic behavior.
First, the position increases up to $z_f=0.4$ at $\Phi=140^{\circ}$, it then subsequently decreases to $z_f\approx0.3$ at $\Phi=200^{\circ}$, and finally increasing once again to $z_f\approx0.4$ at $\Phi=260^{\circ}$. 
For $280^{\circ} \leq \Phi \leq 320^{\circ}$, the eccentric focusing position remains at $z_f \approx 0.4$, but becomes less distinct, while coexisting with an increasingly well-defined focusing position at the channel centerline. 
For $340^{\circ} \leq \Phi \leq 360^{\circ}$, the eccentric focusing band disappears and the system reverts to single band focusing at the channel centerline.

The analogous focusing efficiency $F_{20\%}$ measurements are then plotted against phase difference, as shown in Figure \ref{fig:figfoursoeif}(c). 
Here, $F_{20} < 20\%$ corresponds to a decrease in particle concentration while $F_{20} > 20\%$ corresponds to an increase in particle concentration, both with respect to a homogenous and uniform particle concentration. 
The yellow dashed line is $F_{20\%}=20\%$, which represents the case of homogeneous and uniformly mixed particle distribution, shown in Figure \ref{fig:figthreesoeif}(a-d). 
For $0^{\circ} \leq \Phi \leq 40^{\circ}$, the focusing efficiency at the channel centerline
is up to $90\%$, as shown in Figure \ref{fig:figfoursoeif}(c, light red squares). 
For $60^{\circ} \leq \Phi \leq 80^{\circ}$, the focusing efficiency decreases as the system transitions from single band to two band focusing. 
For $100^{\circ}\leq\Phi\leq240^{\circ}$, the focusing efficiency within the two eccentric focusing bands (light blue squares) increases from $50\%$ at $\Phi=100^{\circ}$, up to a maximum of $85\%$ at $\Phi=200^{\circ}$, and then proceeds to decrease to $70\%$ at $\Phi=240^{\circ}$. 
For $260^{\circ}\leq\Phi\leq360^{\circ}$, the focusing efficiency of the eccentric bands drops rapidly from $60\%$ to $\approx0\%$ at $\Phi=320^{\circ}$.
Meanwhile, as the centerline position becomes the dominant focusing position, the focusing efficiency improves from $35\%$ up to $90\%$.

To demonstrate the effect of electric field oscillatory amplitude, the focusing position and focusing efficiency were measured for the same conditions, but with a lower electric field oscillatory amplitude of $E=25$ kV/m. 
The evolution of both the focusing position and the focusing efficiency with $\Phi$ is qualitatively similar to that of the higher electric field oscillatory amplitude, as shown in Figure \ref{fig:figfoursoeif}(b, dark squares) and Figure \ref{fig:figfoursoeif}(c, dark red and blue squares), respectively.
In fact, the focusing positions are nearly identical between the two cases. 
However, a weaker electric field amplitude results in a lesser value of the focusing efficiency in both the center and eccentric focusing positions. 
For example, at $\Phi=0^{\circ}$, the centerline focusing efficiency is $80\%$ compared to that of $90\%$ for the higher electric field amplitude. 
Similarly, at $\Phi=200^{\circ}$, the focusing efficiency is $60\%$ compared to that of $85\%$ for the higher electric field amplitude.

\subsection{Particle migration velocity}

The particle migration velocity, in contrast to the focusing efficiency, is a more direct measurement of forces on the particle and does not depend on the channel length \cite{Zhang}. 
It provides an opportunity to better characterize the mechanism of particle migration. 
Furthermore, measurement of the particle migration velocity benefits greatly from oscillatory flows due to the much longer ($>100\times$) path distances that can be observed without loss in measurement resolution. 
To obtain the migration velocity measurements, particle tracking velocimetry is performed  1 mm away from the inlet of the channel. 
This yields a spatially homogeneous sampling of particles and minimizes difficulties in tracking that can arise due to dense particle localization.

\begin{figure}
\centering
\includegraphics[trim=0 0 0 0,clip,width=\linewidth]{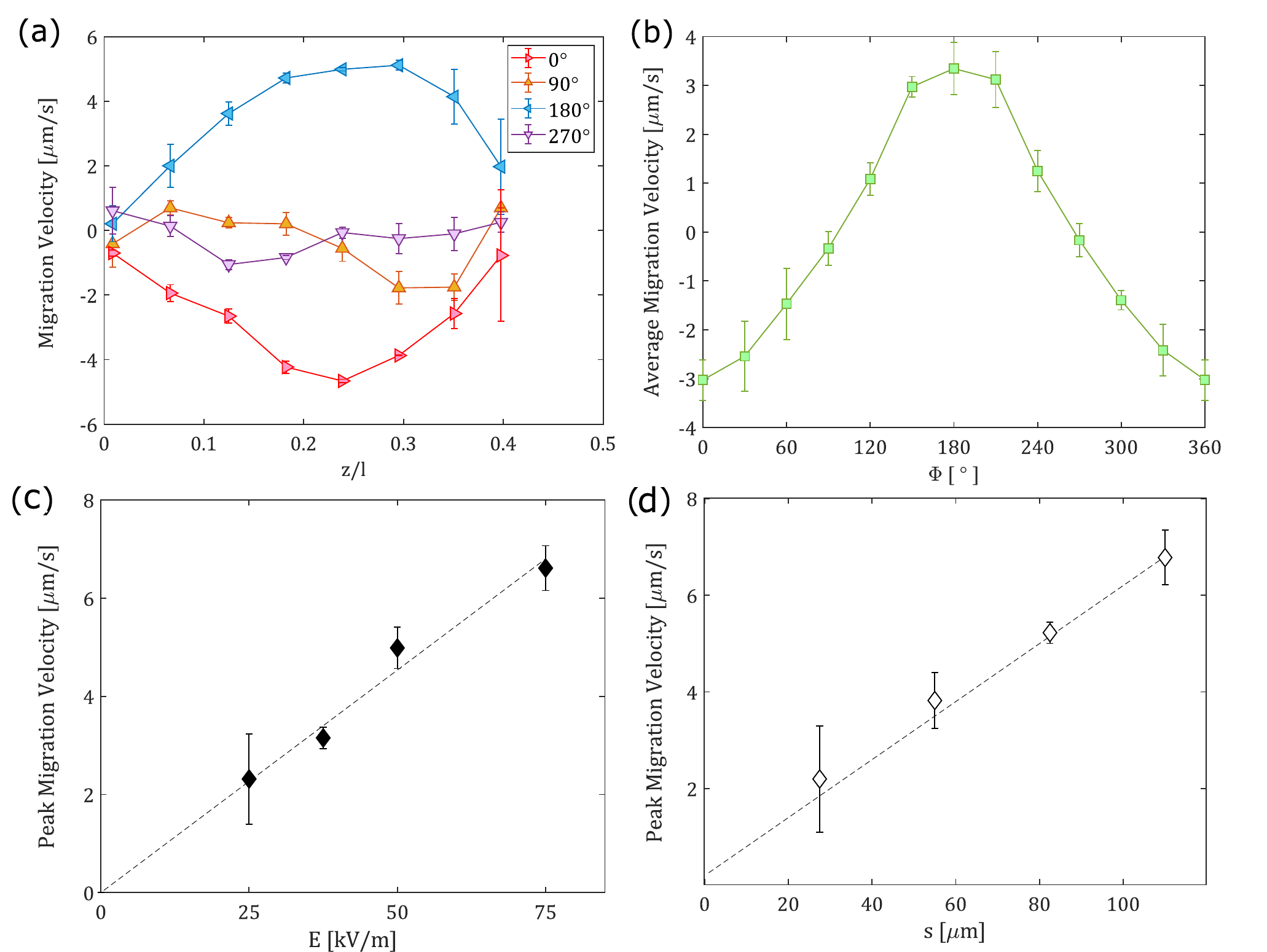}
\caption{
(a) Migration velocity profile for $a=0.5\ \mu$m particles at different values of $\Phi$. 
(b) Spatial average of the migration velocity profile for $a=0.5\ \mu$m particles as a function of the phase difference $\Phi$.
Relevant experimental parameters are: $\bar{u}=1.0$ mm/s, $f=50$ Hz, $E=50$ kV/m, $s=32\ \mu$m, $s_E=7.1\ \mu$m.
(c) Peak migration velocity for $a=0.5\ \mu$m particles as a function of the oscillatory electric field magnitude.
The oscillatory pressure-driven flow amplitude was maintained constant at $s=60.5\ \mu$m. 
(d) Peak migration velocity for $a=0.5\ \mu$m particles as a function of the oscillatory pressure-driven flow amplitude.
The oscillatory electric field magnitude and oscillation amplitude were held constant at $E=50$ kV/m, $s_E=3.4\ \mu$m, respectively. 
}
\label{fig:figfivesoeif}
\end{figure}

The migration velocity profile in the spanwise direction for a particle radius of $a=0.5\ \mu$m was measured for phase differences of $\Phi = 0^\circ, \ 90^\circ, \ 180^\circ, \ 270^\circ$, and is shown in Figure \ref{fig:figfivesoeif}(a).
The profile is extracted by tracking the displacement normal to the flow for each particle per oscillatory period and then taking the median in each segment of the channel width \cite{VishwanathanJFM}. 
For $\Phi=0^{\circ}$, particle migration is directed toward the channel centerline ($z/l=0$) at all positions as indicated by the negative value of migration velocity. 
For $\Phi=90^{\circ}$, the migration velocity magnitude is considerably weaker, but oriented towards the center for $z/l>0.3$ and away from the center for $z/l<0.2$. 
This implies a focusing position located at $z/l \approx 0.22$, which is in agreement with Figure \ref{fig:figfoursoeif}(b). 
For $\Phi=180^{\circ}$, particle migration is directed toward the channel walls; specifically to the focusing position located at $z/l \approx 0.37$. 
It should be noted that the migration velocity profiles of $\Phi=0^{\circ}$ and $\Phi=180^{\circ}$ display mirror symmetry about the $U_m=0$ ordinate. 
Lastly, for $\Phi=270^{\circ}$, the migration velocity magnitude is again weaker and indicates two simultaneous focus positions located at $z/l=0$ and $z/l \approx 0.37$, which is also in agreement with Figure \ref{fig:figfoursoeif}(b). 
In all cases, there is a lack of sufficient particle density very close to the walls (for $|z/l|>0.4$) implying the particle migration velocity cannot be measured accurately, causing larger error.

The spatial average of the migration velocity for different values of the phase difference ranging from $0^{\circ} < \Phi < 360^{\circ}$ was measured, and is shown in Figure \ref{fig:figfivesoeif}(b).
Average migration velocity values appear to follow an approximately $U_m \propto -\cos{\Phi}$ dependence in magnitude.
For $\Phi = 0^{\circ}$ and $360^{\circ}$, both phase differences were observed to produce channel centerline focusing with an average migration velocity of $-3 \ \mu$m/s. 
For $\Phi = 180^{\circ}$, this phase difference produced two band focusing toward the channel walls with an average migration velocity of $3 \ \mu$m/s.
The relatively large values of average migration velocity for these phase differences contribute to high focusing efficiencies, as shown in Figure \ref{fig:figfoursoeif}(c).

Next, the effect of the electric field and oscillatory flow amplitude on the scaling of migration velocity are evaluated. 
The peak migration velocity is plotted against a varying electric field amplitude with the oscillatory flow amplitude and other experimental parameters kept constant, as shown in Figure \ref{fig:figfivesoeif}(c).
The peak value is taken to be the mean of the three largest migration speeds (the local maxima and one point on either side). 
The errorbar corresponds to the difference between the peak values for $z/l>0$ and $z/l<0$. 
The maximum migration velocity scales linearly with electric field amplitude.
Similarly, the peak migration velocity is plotted against the corresponding amplitude of the oscillatory flow with the electric field amplitude and other experimental parameters kept constant, as shown in Figure \ref{fig:figfivesoeif}(d). 
The peak migration velocity also scales linearly with oscillatory flow amplitude.

\subsection{Particle size and migration velocity}

Finally, the particle migration velocity profiles for particles with different radii of $a=1.7 \ \mu$m and $2.6\ \mu$m were measured.
To control for measurement errors of oscillatory flow amplitude and electro-osmotic flow amplitude,
the migration profiles were measured simultaneously by using a dilute solution that contained three different particle sizes. 
Specifically, a solution with a volume fraction of $0.001\%$ of $0.5\ \mu$m particles, a volume fraction of $0.003\%$ of $1.7\ \mu$m particles, and a volume fraction of $0.01\%$ of $2.6\ \mu$m particles was used. 
Larger volume fractions were used for larger particles so that a sufficient count of particle tracks could be obtained for each particle size.

\begin{figure}
\centering
\includegraphics[width=\linewidth]{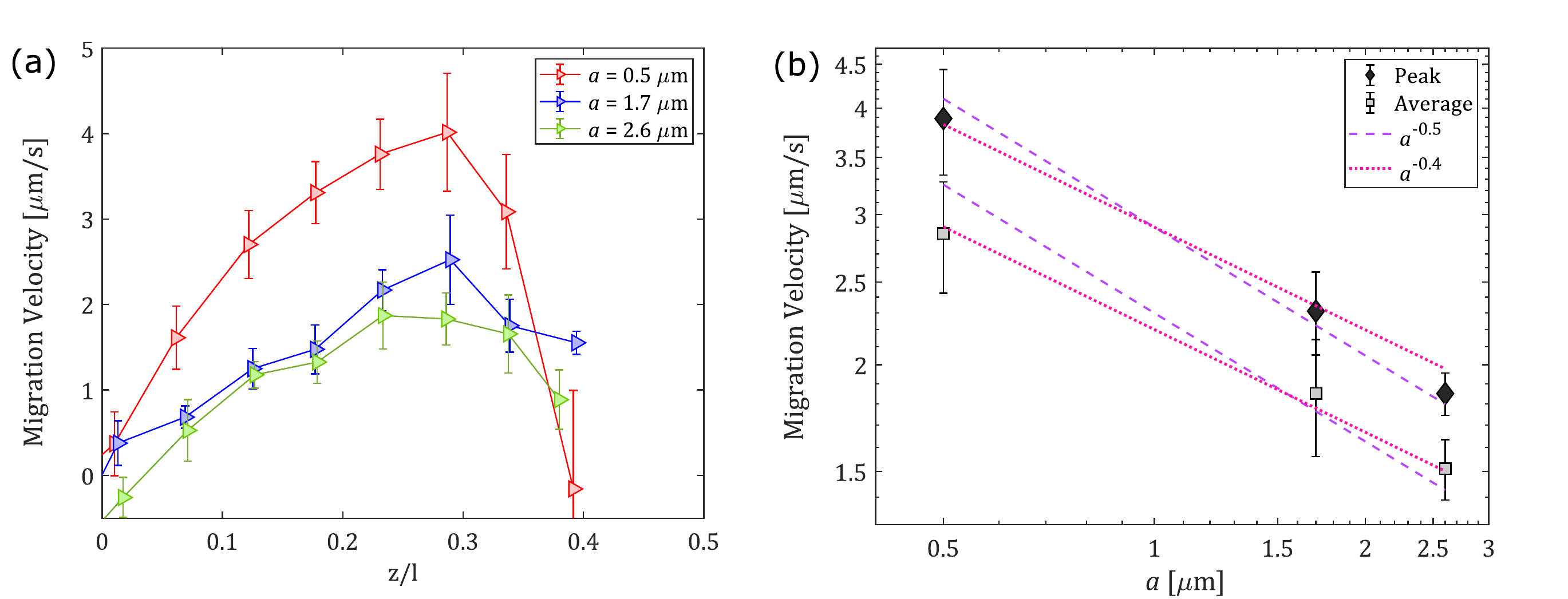}
\caption{
(a) Migration velocity profiles for particles with different radii of $a=0.5,\ 1.7$, and $2.6\ \mu $m. 
Results are obtained for the phase differences of $\Phi=0^{\circ}$ and $\Phi=180^{\circ}$, where the migration speeds are the largest.
(b) Peak and phase averaged migration velocities as a function of particle size. 
Phase averaged velocity is obtained by averaging over $\Phi = 0^{\circ} - 360^{\circ}$ at each $30^{\circ}$ interval.
Dashed and dotted lines are power-law curves for $a^{-0.5}$ and $a^{-0.4}$, respectively. 
Relevant parameters for these experiments are: $\bar{u}=1$ mm/s, $f=50$ Hz, $E=50$ kV/m, $s=32\ \mu$m, $s_E=7.1\ \mu$m.
}
 \label{fig:figsixsoeif}
\end{figure}

The corresponding migration velocity profiles for all three particle sizes are shown in Figure \ref{fig:figsixsoeif}(a).
Here, the migration velocity profiles were computed by measuring the migration velocities at phase differences of $\Phi=0^{\circ}$ and $\Phi=180^{\circ}$ followed by taking the mean of the absolute values.
It is revealed that the particle migration velocity is larger for smaller particles and decreases with particle size provided all other factors are kept identical. 
The migration velocities for other phase differences (not shown) also corroborate this result.

To compare the relation between migration velocity and particle size, the peak and the phase average migration velocity are measured for the three different radii of $a=0.5, 1.7, 2.6\ \mu$m, as shown in Figure \ref{fig:figsixsoeif}(b).
The peak velocity for each particle radius is obtained by taking the mean of the three maximum values for the corresponding curve shown in Figure \ref{fig:figsixsoeif}(a). 
Since the peak velocity represents the maximum migration velocity encountered by a particle it is obtained only from the $\Phi=0^{\circ}$ and $\Phi=180^{\circ}$.
To account for different particle radii possibly having different dependence on phase difference (for instance, attaining maximum values at different phases), the phase average migration velocity is also shown in Figure \ref{fig:figsixsoeif}(b). 
This is obtained by averaging the absolute values of migration velocity profiles for phase $\Phi = 0^{\circ} - 360^{\circ}$ at each $30^{\circ}$ interval and extracting the peak.
While the peak and the phase average values have different magnitudes, the trends are nearly identical. 
The migration velocity decreases with increasing particle size.
For the range of particle radii investigated here, the exponent for the peak and phase average migration velocities appear to follow a particle radius scaling of $\sim a^{-\xi}$, where $0.4 < \xi < 0.5$.

\section{Discussion}

Combining the above results, the following scaling law for the primary mechanism of particle migration in the channel bulk is inferred to be,
\begin{equation} \label{scaling}
    U_m\sim EU\cos{(\Phi+\phi_s)} a^{-\xi}.
\end{equation}
Based on the experimental evidence presented here, the scaling with $\cos{\Phi}$ demonstrated in Figure \ref{fig:figfivesoeif}(b), emerges naturally from having $U_m\propto UE$ where $U$ and $E$ vary sinusoidally in time with the same frequency. 
Furthermore, the linear scaling of $U_m$ with $E$ and $U$ (or $U/l$ for a fixed $l$) are confirmed in Figure \ref{fig:figfivesoeif}(c) and \ref{fig:figfivesoeif}(d), and are in agreement with the results of other authors \cite{Yee2018, Yee2021,Lochab2021}.
There is nevertheless, some indication of a sub-dominant secondary mechanism that causes the evolution of $z_f$ and $F_{20\%}$ with $\Phi$ to be asymmetric about $\Phi=180^{\circ}$, as shown in Figure \ref{fig:figfoursoeif}(b) and \ref{fig:figfoursoeif}(c). 
This fact is reflected in the particle migration velocity profiles shown in Figure \ref{fig:figfivesoeif}(a). 
The migration velocity magnitudes for $\Phi=90^{\circ}$ and $\Phi=270^{\circ}$ are weak, but not uniformly zero within measurement error. 
The negative exponent scaling of $U_m\propto a^{-\xi}$ with $\xi \approx 0.5$ is not conclusive by itself since only 3 different particle radii are considered and the particles have different values of zeta potential $\zeta_p$.
However, it does suggest that $U_m$ scales weaker with particle radius than expected. 



The weak scaling of the magnitude of migration velocity with particle radius is also bolstered by comparison with the experimental results of other authors, summarized in Table \ref{table:comparison}. 
From left to right, the table displays the manuscript reference, the channel geometry, particle radius, characteristic strain rate, the electric field magnitude, characteristic migration velocity and normalized migration velocity rescaled by the strain rate and electric field to account for the established linear scaling. 
It is evident from the table that both the absolute migration velocity magnitude $U_m= 1.5 - 7\ \mu$m/s and the rescaled migration velocity $lU_m/UE = 1.25 - 8 \ \mu$m$^2$/V vary by at most a factor of $6.5$ as the particle radius varies by a factor of 20 from $0.25 \leq a \leq 5\ \mu$m. 
It is noted that for all the works referenced in Table \ref{table:comparison} there are subtle differences in the precise method of migration velocity estimation. 
Nevertheless, they are alike in that they are interested in observing the cross-stream migration of polystyrene microparticles in microchannels in the presence of a simultaneous electric field and fluid flow. 
It is assumed that any differences in the methodology only affect the migration velocity by a near unity factor. 
For our results, the migration velocity is multiplied by a factor of exactly 2 to account for the oscillatory flow and field.

\begin{table}
	\begin{tabular}{lcccccc}
		\hline
		\begin{tabular}[c]{@{}c@{}} Reference \end{tabular} & \begin{tabular}[c]{@{}c@{}} Geometry   \\ {[}$\mathrm{\mu m} ${]}\end{tabular}  &
		\begin{tabular}[c]{@{}c@{}} $a$   \\ {[}$\mathrm{\mu m} ${]}\end{tabular} & 
		\begin{tabular}[c]{@{}c@{}} $U/l$   \\ {[}$\mathrm{s}^{-1}${]}\end{tabular} & 
		\begin{tabular}[c]{@{}c@{}} $E$     \\ {[}kV/m{]}\end{tabular} & 
		\begin{tabular}[c]{@{}c@{}} $U_m$ \\ {[}$\mathrm{\mu m/s} ${]}\end{tabular}  &
		\begin{tabular}[c]{@{}c@{}} $lU_m/EU$   \\ {[}$\mathrm{\mu m^2/V} ${]}\end{tabular}  
		\\ \hline
		Kim \& Yoo (2009) \cite{KimYoo} & Tube $85$  & 2.5 & 20.0 & 12.9  & 1.89 & 7.3  \\ \hline
		Kim \& Yoo (2015) \cite{Yoo}  &  $82\times 90$  & 2.5 & 18.6 & 13.0 & 1.91 & 7.9  \\ \hline
		Li \& Xuan (2018) \cite{Li} & $50 \times 100$ & 5 & 55  & 30 & 6.9 & 4.18   \\ \hline
		Yee \& Yoda (2018) \cite{Yee2018} &  $34.6\times 340$ & 0.25  & 440  & 1 & 1.76 & 4.0  \\ \hline
		Lochab \& Prakash (2021) \cite{Lochab2021} & $34.6\times 340$ & 0.25 & 1000 & 2 & 2.5 & 1.25 \\ \hline
		This work & $300\times1500$ &   0.5  &  36  & 50  & 6 & $6.7^*$   \\ \hline
	\end{tabular}
\caption{ 
Comparison of experimental parameters across studies. 
$^*-$ Multiplied by a factor of two to account for oscillatory flow.
}
\label{table:comparison}
\end{table}

The most promising theoretical prediction of Equation (\ref{scaling}) was derived independently \cite{Choudhary, Kabarowski}. 
For a pure shear flow of a Newtonian liquid with shear rate $u/l$, the migration velocity is given by:
\begin{equation}
U_m=-\frac{0.29\varepsilon \zeta_p a^2EU}{\mu^2 l}.     
\end{equation}
In this case, the lift on the particle arises from a nonlinear interaction between the electrophoretic disturbance velocity created by a charged particle in an electric field, with the undisturbed pressure-driven flow. 
This theory captures many features of the measured electro-inertial migration such as the linear proportionality of both field and amplitude (or equivalently shear rate) and the strong inverse dependence on viscosity \cite{Lochab2021} and predicts the direction of migration correctly. 
For relatively large particles $a>5\ \mu$m, this theory predicts migration velocities comparable in magnitude to experiments. 
For colloidal particles, for instance, $a=500$ nm, the migration velocities are $\sim 10$ nm/s, a hundred times smaller than the experimentally measured migration velocities of $\sim1\ \mu$m/s. 
Furthermore, this theory predicts a strong increase in migration velocity with particle radius, while experiments shown here and elsewhere \cite{Yee2018} suggest a weak and possibly negative scaling with particle radius.

The mechanism responsible for this weak negative scaling with particle radius remains elusive, but it seems plausible that this is caused by a finite value of $\lambda_D$. 
This is suggested by another theory that predicts a Magnus-like lift arising from a shear-induced asymmetry in the ionic cloud surrounding the particle \cite{Balu}. 
The theory predicts a migration velocity:
\begin{equation}
    U_m=-\mathcal{L}(a/\lambda_D)\frac{\varepsilon^2 \zeta_p^3 a^2 EU}{12\pi D^2\mu^2 l},
\end{equation}
where the function $\mathcal{L}(a/\lambda_D)\propto (a/\lambda_D)^{-5}$ when $a\gg\lambda_D$ indicating the possibility of an apparent negative scaling with radius. 
Nevertheless, this theory also predicts migration velocities much lower than those obtained experimentally.

Finally, there is also the possibility of long-range interactions and collective dynamics of particles. 
This was suggested as reasoning for the decrease in accumulation time as the volume fraction increased from $0.17\%$ to $0.33\%$, with all other parameters kept constant \cite{Yee2018}.
In the accumulation time measurements, particles became highly localized near the bottom plane and subsequently self-organized into bands.
Inter-particle interactions are thus inevitable, confounding the relationship between the bulk migration velocity and the accumulation time. 
In this work, however, the effects of particle interactions are significantly reduced by using volume fractions less than $0.001\%$ and measuring migration velocities near the inlet, where particles are homogeneously distributed, and hence, localization and consequent interactions are minimal. 
Nevertheless, comparable magnitudes of migration velocity are obtained, and therefore, collective dynamics and long-range interactions are unlikely.

\section{Conclusions}

In summary, a synchronous oscillatory pressure-driven flow and oscillatory electric field of $f=50$ Hz was used to focus a suspension of polystyrene particles with radii of $0.5 \ \mu$m.
Depending on the phase difference between the oscillatory flow and the electric field, the focusing position, the number of focusing bands, and the focusing efficiency could be controlled. 
Using this experimental setup, migration velocity profiles were accurately measured with average values in the range of $1 < U_m < 10 \ \mu$m/s. 
This measurement was extended to investigate the dependence of the migration velocity on particle size. 
However, there was no conclusive scaling that agrees with the current theories. 
Future experiments could investigate in more detail the dependence on oscillation frequency and particle zeta potential and wall zeta potential.
This is the first experimental attempt at examining the oscillatory analog of the interaction between inertial and electro-kinetic effects at low frequencies. 
Analogous to electrophoresis and dielectrophoresis, at higher frequencies, this electro-inertial coupling could result in entirely new applications. 
The potential advantages of this technique include minimizing electrolysis and electric damage to cells, reduced shear rates, and the ability to manipulate far smaller particles.

\begin{acknowledgments}
We wish to acknowledge Kyle C. Smith, Qian Chen, Sascha Hilgenfeldt, and Aditya S. Khair for the thoughtful comments and constructive criticism that helped improve this manuscript.
\end{acknowledgments}

\section*{Data Availability Statement}
The data that support the findings of this study are available from the corresponding author upon reasonable request.

\bibliography{soeif}

\end{document}